\pdfoutput=1

\documentclass[11pt,twoside,a4paper,cmspaper,final,collab]{cms-tdr}
\def\svnVersion{253136}\def\cmsCernNo{2013-037}\def\cmsCernDate{\today}\def\cmsMessage{Published in Physics Letters B as \href{http://dx.doi.org/10.1016/j.physletb.2012.07.059}{\doi{10.1016/j.physletb.2012.07.059.}}}

\begin{document}\cmsNoteHeader{EXO-11-050}

\hyphenation{had-ron-i-za-tion}
\hyphenation{cal-or-i-me-ter}
\hyphenation{de-vices}

\RCS$Revision: 138777 $
\RCS$HeadURL: svn+ssh://alverson@svn.cern.ch/reps/tdr2/papers/EXO-11-050/trunk/EXO-11-050.tex $
\RCS$Id: EXO-11-050.tex 138777 2012-07-21 01:10:28Z linacre $
\newlength\cmsFigWidth
\newlength\cmsFigNarrowWidth
\ifthenelse{\boolean{cms@external}}{\setlength\cmsFigWidth{0.95\columnwidth}}{\setlength\cmsFigWidth{0.8\textwidth}}
\ifthenelse{\boolean{cms@external}}{\setlength\cmsFigNarrowWidth{0.95\columnwidth}}{\setlength\cmsFigNarrowWidth{0.6\textwidth}}
\cmsNoteHeader{EXO-11-050} 
\title{Search for heavy, top-like quark pair production in the dilepton final state in pp collisions at $\sqrt{s}=7$~TeV}

\date{\today}

\newcommand{\mll}{\ensuremath{m_{\ell\ell}}}
\newcommand{\eepm}{\ensuremath{\Pep\Pem}}
\newcommand{\mmpm}{\ensuremath{\Pgmp\Pgmm}}
\newcommand{\empm}{\ensuremath{\Pe^\pm \Pgm^\mp}}
\newcommand{\ase}[2]{\ensuremath{_{~- #1}^{~+ #2}}}
\newcommand{\met}{\ensuremath{\mspace{3mu}/\mspace{-12.0mu}E_{T}}}
\providecommand{\PYTHIA} {{\textsc{pythia}}\xspace}
\providecommand{\MADGRAPH} {\textsc{MadGraph}\xspace}

\newcommand{\Lumiyt}{5.0\fbinv}
\newcommand{\ttbarit}{$t\bar{t}$}
\newcommand{\ttbarprime}{\ensuremath{\cmsSymbolFace{t'}\overline{\cmsSymbolFace{t'}}}\xspace}
\newcommand{\massltb}{$M_{\mathrm{l1b1}}$}
\newcommand{\massllb}{$M_{\mathrm{l2b2}}$}
\newcommand{\singlet}{\ensuremath{\cmsSymbolFace{t}}\xspace}
\newcommand{\tprime}{\ensuremath{\cmsSymbolFace{t'}}\xspace}
\newcommand{\bprime}{\ensuremath{\cmsSymbolFace{b'}}\xspace}
\newcommand{\tbar}{\ensuremath{\overline{\cmsSymbolFace{t}}}\xspace}
\newcommand{\tprimebar}{\ensuremath{\overline{\cmsSymbolFace{t'}}}\xspace}
\newcommand{\tprimemass}{\ensuremath{M_{\cmsSymbolFace{t'}}}\xspace}
\newcommand{\bprimemass}{\ensuremath{M_{\cmsSymbolFace{b'}}}\xspace}
\newcommand{\tmass}{$M_{\mathrm{t}}$}
\newcommand{\ttll}{$\ensuremath{\cmsSymbolFace{t}\overline{\cmsSymbolFace{t}}}\xspace\rightarrow\ell^+\ell^-$}
\newcommand{\tttau}{$\ensuremath{\cmsSymbolFace{t}\overline{\cmsSymbolFace{t}}}\xspace\rightarrow\ell^{\pm}\tau^{\mp}/\tau^+\tau^-$}
\newcommand{\ttfake}{$\ensuremath{\cmsSymbolFace{t}\overline{\cmsSymbolFace{t}}}\xspace\rightarrow\mathrm{other}$}
\newcommand{\wjets}{$\PW+\mathrm{jets}$}
\newcommand{\DY}{DY}
\newcommand{\DYll}{DY$\rightarrow\ell^+\ell^-$}
\newcommand{\WW}{$\PWp\PWm$}
\newcommand{\WZ}{$\PW^{\pm}\PZ$}
\newcommand{\ZZ}{$\PZ\PZ$}

\newcommand{\Ht}{$H_\mathrm{T}$}
\newcommand{\ptll} {\ensuremath{\pt(\ell\ell)}}
\newcommand{\ptht} {\ensuremath{\pt(\ell\ell/\sqrt{\Ht}}}
\newcommand{\RelIso}{\ensuremath{I_\text{rel}}}
\newcommand{\masslbMin}{\ensuremath{M_{\ell\cPqb}^{\text{min}}}\xspace}
\newcommand{\masslb}{\ensuremath{M_{{\ell\cPqb}}}\xspace}

\abstract{

The results of a search for pair production of a heavy, top-like quark, \tprime, in the decay mode
$\ttbarprime \to \cPqb\PWp \cPaqb\PWm \to \cPqb \ell^{+} \cPgn \: \cPaqb \ell^{-} \cPagn$ are presented.
The search is performed with a data sample corresponding to an integrated luminosity of 5.0\fbinv in
\Pp\Pp\ collisions at a center-of-mass energy of 7\TeV, collected by the CMS experiment at the LHC.
The observed number of events agrees with the expectation from standard model processes, and no evidence of
\ttbarprime\ production is found. Upper limits on the production cross section as a function of
\tprime\ mass are presented, and \tprime\ masses below 557\GeVcc are excluded at the 95\% confidence level.
}
\hypersetup{%
pdfauthor={CMS Collaboration},%
pdftitle={Search for heavy, top-like quark pair production in the dilepton final state in pp collisions at sqrt(s) = 7 TeV},%
pdfsubject={CMS},%
pdfkeywords={CMS, physics, dilepton, top-like, tprime, fourth generation}}

\maketitle 

\section{Introduction}
\label{sec:intro}

Since the discovery of the top quark at the Tevatron~\cite{PhysRevLett.74.2626, PhysRevLett.74.2632}, there have been many searches for a possible new generation of fermions.
Those searches have found no evidence of new fermions beyond the standard model (SM). However, based on present knowledge,
there is no compelling reason for the number of fermion generations to be limited to three~\cite{ref:review}. Additional generations of fermions
may have a significant effect on neutrino, flavor, and Higgs physics. A fourth generation of quarks, \tprime\ and \bprime,
may result in enough intrinsic matter
and anti-matter asymmetry to explain the baryon asymmetry of the universe~\cite{ref:theory2}.
Therefore, there is continued
theoretical and experimental interest in the search for a fourth generation fermion~\cite{ref:review}.

Previous direct searches
restrict the masses of quarks in the fourth generation, \tprimemass\ and \bprimemass, to be greater than
404 and 372\GeVcc, respectively, at the 95\% confidence level~\cite{Aad:2012xc, ref:CDFbprime}, and the measurement of the $\cPZ$ lineshape at the Large Electron-Positron
collider excludes a fourth generation of light neutrinos~\cite{ref:lep, Adeva:1989mn, Akrawy:1989pi, Aarnio:1989tv}.
At the Large Hadron Collider (LHC), the quantum chromodynamics (QCD) production cross section of \ttbarprime\ is expected to be
two orders of magnitude larger than at the Tevatron for $\tprimemass =500$\GeVcc~\cite{Berger:2009qy}. This increase provides an
opportunity to explore the possibility of new physics with an additional generation of fermions at higher masses.

We present a search for pair production of a heavy top-like quark in the final state
$\ttbarprime \to \cPqb\PWp \cPaqb\PWm \to \cPqb \ell^{+} \cPgn \: \cPaqb \ell^{-} \cPagn$,
where a branching fraction of 100\% to $\cPqb\PW$ is assumed and the charged lepton is either an electron or a muon.
This search is motivated if $\tprimemass < \bprimemass + M_{\PW}$, which is favored by precision electroweak constraints~\cite{TimTait, Eberhardt}.
The presence of two leptons (dileptons) in the final state helps to suppress SM backgrounds,
providing a clean environment to search for physics beyond the SM.
The data sample corresponds to an integrated luminosity of \Lumiyt\
in pp collisions at a center-of-mass energy of 7\TeV,
collected by the Compact Muon Solenoid (CMS) experiment at the LHC during 2011.

\section{CMS detector}

The central feature of the CMS apparatus is a superconducting
solenoid, 13\unit{m} in length and 6\unit{m} in diameter, which provides
an axial magnetic field of 3.8\unit{T}. Within the field volume are
several particle detection systems.
Charged particle
trajectories are measured by silicon pixel and silicon strip trackers,
covering $0 \leq \phi < 2\pi$ in azimuth and $|\eta| < 2.5$ in pseudorapidity,
where $\eta = -\ln [\tan (\theta/2)]$ and $\theta$ is the
polar angle of the trajectory of the particle with respect to
the counterclockwise proton beam direction. A lead tungstate crystal electromagnetic calorimeter
and a brass/scintillator hadron calorimeter surround the
tracking volume, providing energy  measurements of electrons and
hadronic jets. Muons are identified and measured in gas-ionization detectors embedded in
the steel flux return yoke of the solenoid. The CMS detector is nearly
hermetic, allowing momentum balance measurements in the plane
transverse to the beam direction. A two-tier trigger system selects pp collision events
of interest for use in physics analyses.
A more detailed description of the CMS detector can be found
elsewhere~\cite{JINST}.

\section{Event samples, reconstruction, and preselection}
\label{sec:eventSel}
The data used for this measurement were collected using one of the ee, e$\mu$, or $\mu\mu$
high-\pt\ double-lepton triggers. Muon candidates are reconstructed using two algorithms that require consistent signals in the
tracker and muon systems: one matches the extrapolated trajectories from the silicon tracker to signals in the muon system (tracker-based muons), and
the second performs a global fit requiring consistent patterns in the tracker and the muon system (globally fitted muons)~\cite{MUOPAS}.
Electron candidates are reconstructed starting from a
cluster of energy deposits in the
electromagnetic calorimeter. The cluster is then matched to signals in the silicon tracker.
A selection using electron identification variables based on shower
shape and track-cluster matching is applied to the reconstructed
candidates~\cite{EGMPAS}.
Electron candidates within ${ \Delta R}\equiv\sqrt{{ (\Delta\eta)^2+(\Delta\phi)^2}}<$~0.1
from a muon are rejected to remove candidates due to muon bremsstrahlung and final-state radiation.
Both electrons and muons are required to be isolated from other
activity in the event.
This is achieved by imposing a maximum allowed value of 0.15 on the ratio of the
scalar sum of track transverse momenta and calorimeter transverse energy deposits
within a cone of $\Delta R <$~0.3 around the lepton candidate direction at the origin (the transverse momentum of the candidate is excluded),
to the transverse momentum of the candidate.

Event preselection is applied to reject events other than those from
\ttbar\ or \ttbarprime\ in the dilepton final state.
Events are required to have two opposite-sign, isolated leptons ($\mathrm{e^+e^-}$,
$\mathrm{e^{\pm}}\mu^{\mp}$, or $\mu^+\mu^-$). Both leptons must have transverse momentum $\pt >  20\GeVc$, and the electrons (muons) must have $|\eta| < 2.5$ ($2.4$).
The reconstructed lepton trajectories must be consistent with a common interaction vertex.
In the rare case ($<0.1\%$) of events
with more  than two such leptons,  the two leptons with the
highest \pt are selected.  Events with an $\mathrm{e^+e^-}$  or $\mu^+\mu^-$ pair
with  invariant mass  between 76 and  106\GeVcc or  below
12\GeVcc are removed to suppress Drell--Yan (DY) events ($\mathrm{Z}/\gamma^{*}\to\ell^{+}\ell^{-}$)  as  well  as   low  mass  dilepton
resonances. The jets and the missing transverse energy \MET\ are reconstructed with a particle-flow
technique~\cite{CMS-PAS-PFT-10-002}. The  anti-$k_{\mathrm T}$   clustering
algorithm~\cite{antikt} with a distance parameter of 0.5 is used for jet
clustering. At least two jets with  $\pt > 30\GeVc$ and  $|\eta| < 2.5$,
separated  by $\Delta  R  >$  0.4 from  leptons  passing the  analysis
selection, are required in each event. Exactly two of
these jets are required to be consistent with coming from the decay of heavy flavor hadrons and be identified as b~jets by  the TCHEM b-tagging algorithm~\cite{ref:btag}, which relies on tracks with large impact parameters. The  \MET\ in  the event is required to exceed  50\GeV, consistent with the presence of two undetected neutrinos with large $\pt$.

Signal and background events are generated using the
\MADGRAPH4.4.12~\cite{Madgraph} and \PYTHIA6.4.22~\cite{Pythia}  event
generators. The samples of \ttbar, \wjets, DY with $M_{\ell\ell}>50$\GeVcc, diboson (WW, WZ, and ZZ only: the contribution from W$\gamma$ is assumed to be negligible), and single top quark events are generated using \MADGRAPH.
The DY event samples with $M_{\ell\ell}<50$\GeVcc are generated using \PYTHIA.
The samples of \ttbarprime\ events are generated using \MADGRAPH, but decayed using \PYTHIA.
The $\tprime \to \mathrm{W b}$ decay is modeled assuming a V-A structure of the interaction.
Events are then simulated  using a \textsc{Geant4}-based
model~\cite{Geant} of the CMS  detector, and finally reconstructed and
analyzed with the same software used to process collision data.
The cross section for \ttbar\ production is taken from a recent CMS measurement~\cite{PhysRevD.84.092004}, while next-to-leading order (NLO) cross sections are used for the remaining SM background samples.
The \ttbarprime\ cross sections are calculated to approximate next-to-NLO (NNLO) using \textsc{hathor}~\cite{ref:NNLOxsec}.

With the steadily increasing LHC instantaneous luminosity, the mean
number of interactions in a single bunch crossing also increased over the course of data taking, reaching about 15
at the end of the 2011 running period. In the following, the yields of simulated events are weighted such that the distribution of
reconstructed vertices observed in data is reproduced.
The average efficiency for events containing two
leptons satisfying the analysis selection to  pass at  least  one of the double-lepton
triggers  is measured to be approximately 100\%, 95\%, and 90\%
for the ee, e$\mu$, and $\mu\mu$ triggers, respectively, and corresponding weights are applied to the simulated event yields.
In addition,
b-tagging scale factors are applied to simulated events for each jet,
to account for the difference between b-tagging efficiencies in data and simulation~\cite{ref:btag}.

The observed and simulated yields after the above event preselection are listed in Table~\ref{tab:yields1},
in which the categories \ttll\ and \DYll\ correspond to dileptonic \ttbar\ and DY decays, including $\tau$ leptons only when they also decay leptonically. All
other \ttbar\ decay modes are included in the category \ttfake.  The yields are dominated
by top-pair production in the dilepton final state, and agreement is observed between data and simulation.
The expected yields from \ttbarprime\ are also shown for different values of \tprimemass.

\begin{table*}[htb]
\begin{center}
\caption{\label{tab:yields1}
The observed and simulated yields after the preselection described in the text.
The uncertainties on the yields of the simulated samples are statistical only,
while for the simulated total background yields the systematic uncertainties from the sources described in Section~\ref{sec:systematics} are also given.
For \wjets, where the simulated yields are zero, upper limits are given based on the weighted yield, had one of the simulated events passed the preselection.
}
\setlength{\extrarowheight}{1.5pt}
\begin{tabular}{l |  c  c  c  c}
\hline
Sample  & ee & $\mu\mu$ & e$\mu$ & all \\ \hline
\ttbarprime, \tprimemass\ $= 400$\GeVcc\    &   10.6 $\pm$    0.9 &   13.9 $\pm$    1.0 &   29.4 $\pm$    1.5 &   53.9 $\pm$    2.0 \\
\ttbarprime, \tprimemass\ $= 500$\GeVcc\    &   3.0 $\pm$    0.2 &    3.3 $\pm$    0.2 &    6.7 $\pm$    0.4 &   12.9 $\pm$    0.5 \\
\ttbarprime, \tprimemass\ $= 600$\GeVcc\   &   0.9 $\pm$    0.1 &    1.0 $\pm$    0.1 &    2.2 $\pm$    0.1 &    4.1 $\pm$    0.2 \\  \hline
    \ttll\       & 488 $\pm$   11 &  615 $\pm$   12 & 1472 $\pm$   19 & 2575 $\pm$   25 \\
   \ttfake\     &   7.2 $\pm$    1.3 &    0.5 $\pm$    0.3 &   10.5 $\pm$    1.6 &   18.2 $\pm$    2.1 \\
    \wjets\  &   $<1.9$ &  $<1.9$ & $<1.9$ & $<1.9$ \\
    \DYll\   &   2.9 $\pm$    1.5 &    1.6 $\pm$    1.0 &    0.6 $\pm$    0.5 &    5.1 $\pm$    1.8 \\
       Diboson  &   0.5 $\pm$    0.1 &    1.1 $\pm$    0.2 &    1.9 $\pm$    0.2 &    3.6 $\pm$    0.3 \\
Single top quark &  15.6 $\pm$    1.0 &   19.5 $\pm$    1.1 &   46.9 $\pm$    1.7 &   82.0 $\pm$    2.2 \\   \hline
 Total background  & 514 $\pm$   11 $\pm$ 54 &  637 $\pm$ 12 $\pm$ 67 & 1532 $\pm$ 19 $\pm$ 162 & 2683 $\pm$ 25 $\pm$ 284 \\  \hline
     Data  & 510 & 615 & 1487 & 2612 \\  \hline
\end{tabular}
\end{center}
\end{table*}

\section{Signal region}
After preselection, the sample is dominated by SM \ttbar\ events. Since a \tprime\ quark is expected to have a much larger mass than that of the top quark, variables that are correlated with the decaying quark mass can help distinguish \ttbarprime\ events from \ttbar\ events.
The mass of the system defined by the lepton and b~jet (\masslb) from the quark decay is chosen for this purpose.
In the decay of a given top quark, \masslb\ is less than $\sqrt{{M_{\mathrm{t}}}^{2}-{M_{\mathrm{W}}}^{2}}$, where $M_{\mathrm{t}}$ and $M_{\mathrm{W}}$ are the masses of the top quark and W boson. In contrast, most \tprime\ decays have \masslb\  larger than that value. At the reconstruction level, however, there are two ways to combine the two leptons and two b~jets in each event, giving four possible values of \masslb. The minimum value of the four masses (\masslbMin) is found to be a good variable for distinguishing \ttbarprime\ events from \ttbar\ events.
A comparison between \ttbarprime\ events and \ttbar\ events for this variable is shown in Fig.~\ref{fig:mass1d}.

The signal region is defined by adding the requirement for the minimum mass of lepton and jet pairs to be  $\masslbMin>170$\GeVcc. This additional selection reduces the expected number of \ttbar\ events by four orders of magnitude compared with the preselection prediction given in Table~\ref{tab:yields1}. The simulated yields of \ttbarprime\ events are typically reduced by 50\%; they are given for different values of \tprimemass\ in Table~\ref{table:signal_MC}.

\begin{table*}[hbt]
\begin{center}
\caption{\label{table:signal_MC} The expected yields of \ttbarprime\ events in the signal region for different values of \tprimemass.
Uncertainties are statistical only. 
}
\setlength{\extrarowheight}{1.5pt}
\begin{tabular}{l |  c  c  c  c}
\hline
\ttbarprime\ sample  & ee & $\mu\mu$ & e$\mu$ & all \\ \hline
\tprimemass\ $= 400$\GeVcc\   &   3.5 $\pm$    0.5 &    5.5 $\pm$    0.6 &   11.2 $\pm$    0.9 &   20.1 $\pm$    1.2 \\
\tprimemass\ $= 500$\GeVcc\   &   1.4 $\pm$    0.2 &    1.9 $\pm$    0.2 &    3.3 $\pm$    0.2 &    6.7 $\pm$    0.4 \\
\tprimemass\ $= 600$\GeVcc\   &   0.6 $\pm$    0.1 &    0.6 $\pm$    0.1 &    1.3 $\pm$    0.1 &    2.5 $\pm$    0.1 \\  \hline

\end{tabular}
\end{center}
\end{table*}

\section{Background estimation}
\label{Sec:BkgEst}

One of the main sources of background events in the signal region is the
misidentification of b~jets and leptons. A misidentified lepton is defined as a lepton candidate
not originating from a prompt decay, such as a lepton from a semileptonic b or c quark decay, a
muon from a pion or kaon decay, an unidentified photon conversion, or a pion misidentified as an electron.
Misidentified b~jets are referred to as ``mistags'', and occur when a non-b jet
satisfies the b-tagging requirements.

The background events in the signal region can be divided into the following categories:
\begin{itemize}
              \item Category I: events with mistagged b~jet(s) and two real leptons;
              \item Category II: events with misidentified lepton(s) and two real b~jets;
              \item Category III: events with two real b~jets and two real leptons;
              \item Category IV: events with mistagged b~jet(s) and misidentified lepton(s).
\end{itemize}
For each category, an estimate of the combined yield of ee, e$\mu$, and $\mu\mu$ events is made.

To predict the number of events with mistagged b~jet(s) (Category I), control regions
in data are used where events pass all selection requirements except the number of b-tagged jets.
The number of background events with one mistag, $N_{\mathrm{1 mistag}}$, is estimated from
events with one b~tag. Each event is weighted based on the mistag rate $r_i$ for
each untagged jet in the event, where $r_i$ gives the \pt- and $\eta$-dependent probability (with a mean of 0.02) for a non-b jet to be
b-tagged~\cite{ref:btag}. Where there are no untagged jets passing the \masslbMin\ selection, the event weight is zero, and
for each untagged jet $i$ passing the selection the event weight is increased by $r_i/(1-r_i)$.
The subtraction of $r_i$ in the denominator is necessary to account for non-b jets that were mistagged, and are thus missing from the sample of untagged jets.
A similar calculation is made using events with no b~tags to estimate the number of events
with two mistags, $N_{\mathrm{2 mistags}}$. This time a weight of
$r_i/(1-r_i) \times r_j/(1-r_j)$
is used for each pair of untagged jets passing selection, where $r_i$ and $r_j$ are the mistag rates
for the two untagged jets. The final prediction is obtained as
$N_{\mathrm{mistags}} = N_{\mathrm{1 mistag}} -  N_{\mathrm{2 mistags}}$, which takes into
account that $N_{\mathrm{2 mistags}}$ is counted twice in $N_{\mathrm{1 mistag}}$. The
performance of the method is checked using simulated events, and an under-prediction of up
to 50\% is observed. We therefore assign a large systematic uncertainty, 100\%, to this prediction.
In data, the predicted number of events with mistags in the signal region is
$N_{\mathrm{mistags}} = 0.7 \pm 0.3 \pm 0.7$, where the
uncertainties are statistical and systematic, respectively.
The Category I yield in the simulation, taken as a cross-check using the samples mentioned in Section~\ref{sec:eventSel}, is
$1.0 \pm 0.3$,
and is consistent with the prediction based on data.

The background from events with misidentified leptons
(Category II) is predicted based on the number of events in data with a candidate lepton that can
pass only loosened selection criteria~\cite{ref:samesign}. Using a measurement of the
fraction of such ``loose'' leptons that go on to pass the selection requirements, the number of misidentified leptons in the event sample can be estimated. However, there are no observed data events where
one or more of the lepton candidates passes only the loosened selection criteria, resulting
in a prediction of $0.0^{+0.4}_{-0.0}$ events where the upper uncertainty corresponds to the prediction of the method,
had there been one such event. The Category II event yield is also zero in the simulation.

The simulation is used to predict the number of events with no misidentified b~jets or leptons (Category
III), using the background event samples of Section~\ref{sec:eventSel}. Selecting only events where both b~jets and leptons are well matched to the corresponding
particles at the generator level, the resulting prediction is
$1.0 \pm 0.7$
where the
uncertainty is statistical. The systematic uncertainty is small in comparison (Section~\ref{sec:systematics}), so the total uncertainty is also 0.7.

The contribution of events from Category IV is found to be negligible and is covered by both
the Category I and Category II predictions. Since the Category II prediction is zero,
there is no possibility of double-counting.

\section{Systematic uncertainties}
\label{sec:systematics}
The systematic uncertainty on the overall selection efficiency is dominated by the uncertainty on the b-tagging efficiency. This uncertainty is 15\% for b~jets with $\pt > 240\GeVc$, and 4\% for b~jets with $\pt \le 240\GeVc$~\cite{ref:btag}.
Other uncertainties include those on trigger efficiency (2\%), lepton selection (2\%), and jet and \MET\ energy scale (8\%)~\cite{ref:jes}. These four sources combine to yield a 19\% relative uncertainty on the overall selection efficiency for signal events. There is a further 2.2\% uncertainty on the luminosity measurement~\cite{ref:lumi}.

The systematic uncertainty on the background estimate is dominated by the uncertainty on
the estimate of events with mistagged b~jets from data (100\%), and
by the lack of selected events in the loose-lepton control region. The systematic uncertainties on these sources of background are included in the summary of background predictions given in Table~\ref{tab:results}.

\section{Results and summary}
\label{sec:results}

The number of expected events from background processes is $1.8 \pm 1.1$,
and one event is observed in the e$\mu$ channel. There is thus no evidence for an excess of events above
SM expectations. A summary of the observed and predicted yields is presented in Table~\ref{tab:results}.

\begin{figure}[hbt]
\begin{center}
\includegraphics[width=\cmsFigNarrowWidth]{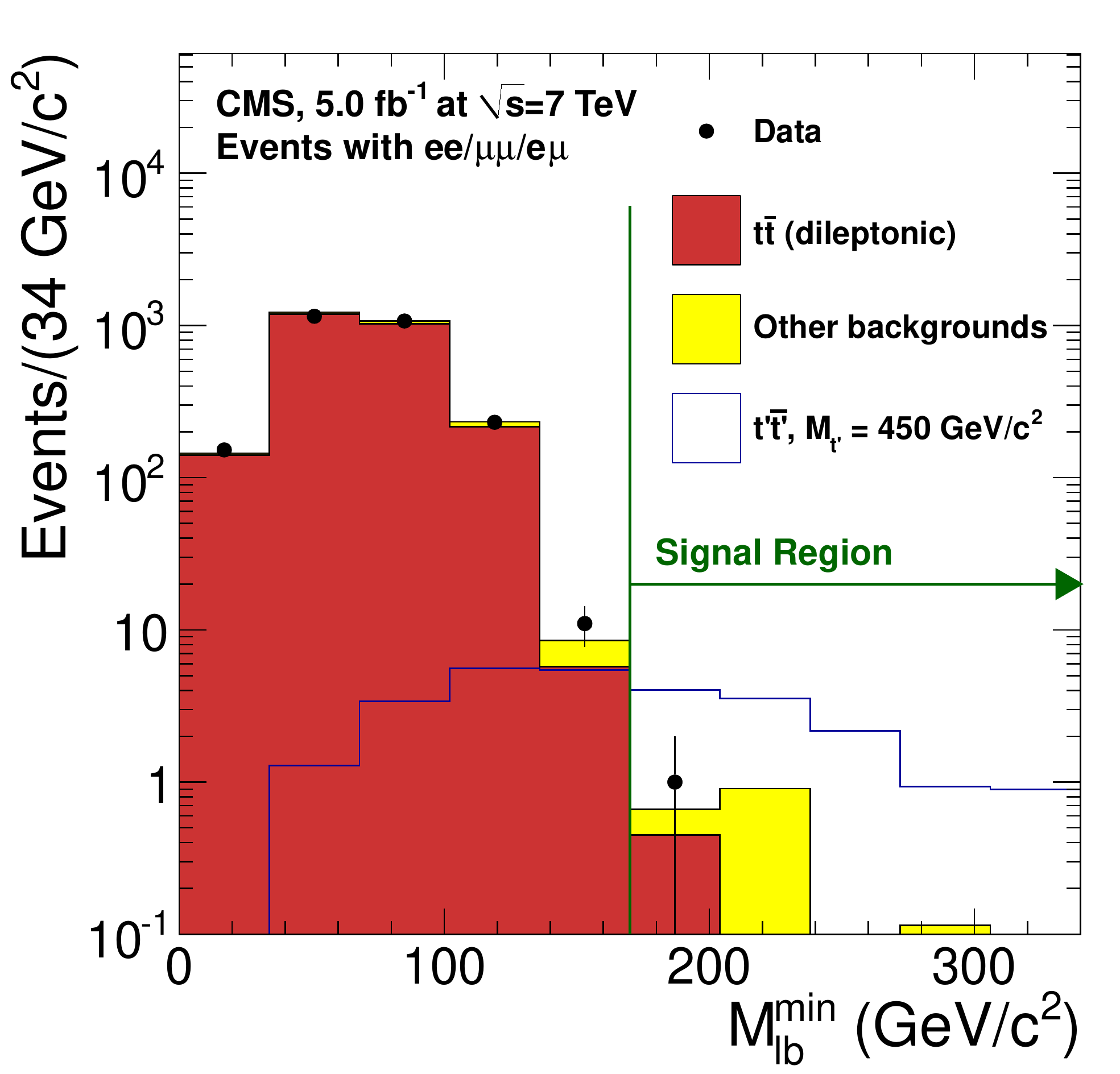}
\caption{\label{fig:mass1d}\protect
Comparison between the data and the simulated background for \masslbMin.
The signal region is defined by \masslbMin~$>170$\GeVcc.
The Category I simulated background yield in the signal region is scaled so that it matches the yield estimated from control regions in data, as given in Table~\ref{tab:results}.
Outside the signal region the simulated background yields are taken without rescaling.
One event is observed in the signal region. The expected distribution for a \ttbarprime\ signal with \tprimemass~$=450$\GeVcc is also shown.
}
\end{center}
\end{figure}

The simulated distribution of \masslbMin\ from background processes is compared with the data in Fig.~\ref{fig:mass1d},
where the expected distribution for a \ttbarprime\ signal
with $\tprimemass=450\GeVcc$ is also shown.

\begin{table}[hbt]
\begin{center}
\caption{\label{tab:results}
Summary of the predicted background yields and the measured yield in data. Statistical and systematic uncertainties are combined.
}
\begin{tabular}{l |  c }
\hline
Sample  & Yield \\ \hline
Category I  (from data) &   $0.7 \pm 0.8$ \\
Category II (from data) &  $ 0.0^{+0.4}_{-0.0}$ \\
Category III (simulated) & $1.0 \pm 0.7$ \\ \hline
Total prediction &    $1.8 \pm 1.1$ \\  \hline
Data  & 1  \\  \hline
\end{tabular}
\end{center}
\end{table}

Finally, 95\% confidence level (CL) upper limits on the production cross section of \ttbarprime\  as a function of \tprime\ mass are set, using the CL$\mathrm{_s}$ method~\cite{ref:cls1, ref:cls2},
where nuisance parameters are varied in the ensemble tests using log-normal distributions.

\begin{figure}[!htpb]
\begin{center}
\includegraphics[width=\cmsFigWidth]{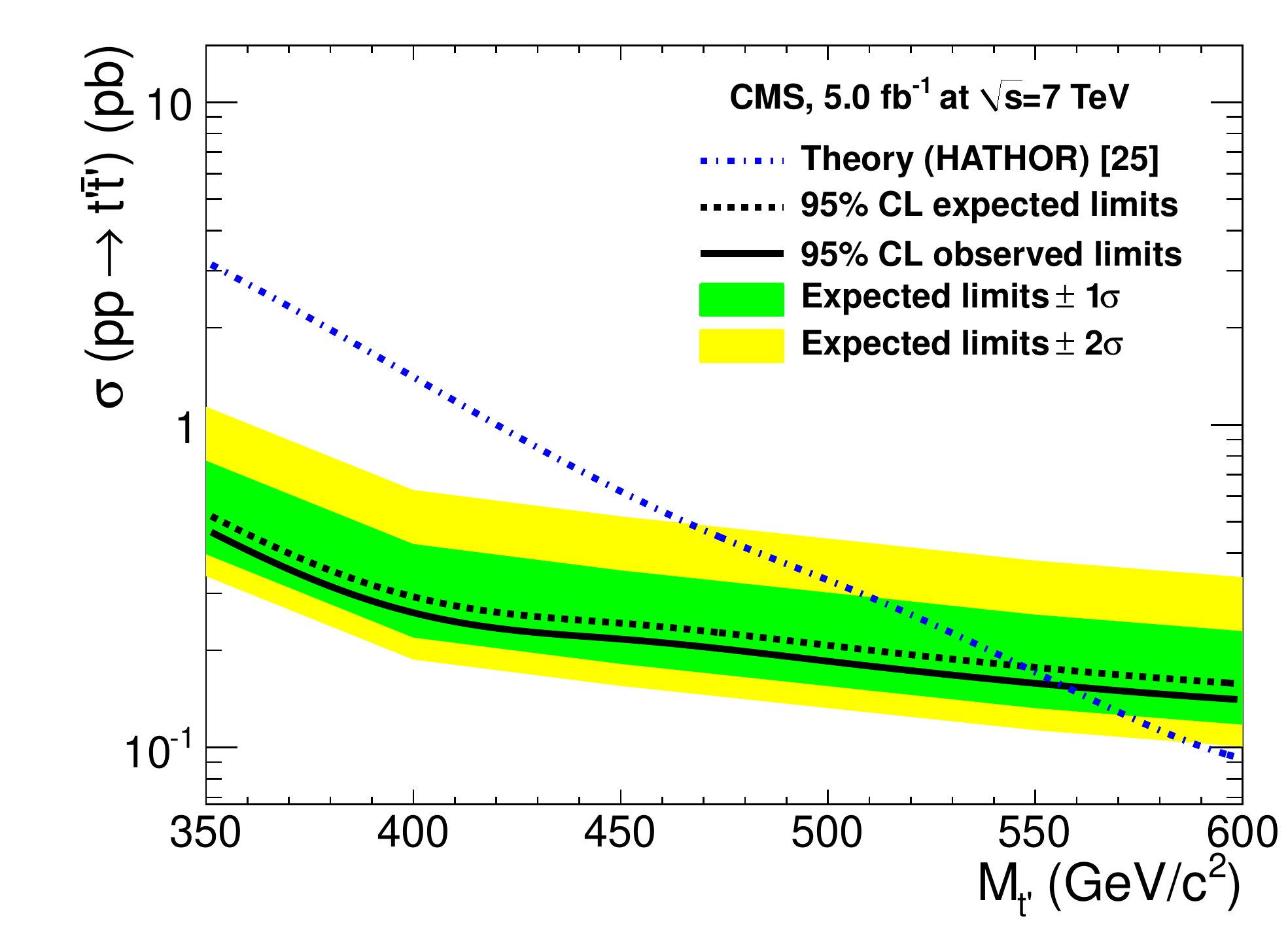}
\caption{\label{fig:limit} The 95\%~CL upper limits on the production cross section of \ttbarprime\  as a function of \tprime\ mass. The observed (expected) 95\%~CL lower bound on \tprimemass\  is 557 (547) \GeVcc. }
\end{center}
\end{figure}

The limit calculation is based on the information provided by the observed event count combined with the values and the uncertainties of the luminosity measurement, the background prediction, and the fraction of \ttbarprime\ events expected to be selected.
This fraction (the overall selection efficiency) is taken as the product of efficiency, acceptance, and the branching fraction for simulated signal events, and is given in Table~\ref{tab:effAccBR} for different values of \tprimemass. The calculated limits are shown in Table~\ref{tab:limit} and Fig.~\ref{fig:limit}.

In summary, assuming a branching fraction of 100\% for $\tprime \to \mathrm{bW}$,
the expected and observed 95\%~CL lower bounds on the \tprime\ mass are $547$ and $557$\GeVcc, respectively, from the analysis of a data sample of pp collisions at $\sqrt{s}=7$\TeV, corresponding to an integrated luminosity of \Lumiyt.

\begin{table}[!htpb]
\begin{center}
\caption{\label{tab:effAccBR} Overall selection efficiency in simulated events for different \tprime\ masses. The branching fraction of
6.5\% for the dilepton decay mode of \ttbarprime\ is included.
The uncertainties are calculated using the systematic uncertainty of 19\% from Section~\ref{sec:systematics}.
}
\setlength{\extrarowheight}{1.5pt}
\begin{tabular}{l | c }
\hline
\ttbarprime sample  & Eff$\times$Acc$\times$Br (\%) \\ \hline
 \tprimemass\ $= 350$\GeVcc\   &   0.16 $\pm$ 0.03\\
 \tprimemass\ $= 400$\GeVcc\   &   0.29 $\pm$ 0.06\\
 \tprimemass\ $= 450$\GeVcc\   &   0.35 $\pm$ 0.07\\
 \tprimemass\ $= 500$\GeVcc\   &   0.41 $\pm$ 0.08\\
 \tprimemass\ $= 550$\GeVcc\   &   0.48 $\pm$ 0.09\\
 \tprimemass\ $= 600$\GeVcc\   &   0.54 $\pm$ 0.10\\  \hline
\end{tabular}
\end{center}
\end{table}

\begin{table*}[!htpb]
\begin{center}
\caption{\label{tab:limit} The approximate NNLO theoretical cross section
of \ttbarprime\ production assuming standard QCD couplings~\cite{ref:NNLOxsec}, and the expected and observed 95\%~CL upper limits on the production cross section of \ttbarprime, for different \tprime\ masses.  }
\setlength{\extrarowheight}{1.5pt}
\begin{tabular}{lcccccc}
\hline
  \tprimemass\ (\GeVcc) & 350 & 400 & 450    & 500  & 550  & 600  \\
\hline
    Theoretical cross section (pb) &  3.20  &  1.41 & 0.62  & 0.33 & 0.17 & 0.09 \\
\hline
    Expected limit (pb) & 0.53 & 0.29 & 0.24 & 0.21 & 0.18 & 0.16 \\
\hline
    Observed limit (pb) & 0.47 & 0.26 & 0.22 & 0.18 & 0.16 & 0.14 \\
\hline
  \end{tabular}
\end{center}
\end{table*}

\section*{Acknowledgment}
We congratulate our colleagues in the CERN accelerator departments for the excellent performance of the LHC machine. We thank the technical and administrative staff at CERN and other
CMS institutes, and acknowledge support from: FMSR (Austria); FNRS and FWO (Belgium); CNPq, CAPES, FAPERJ, and FAPESP (Brazil); MES (Bulgaria); CERN; CAS, MoST, and NSFC (China);
COLCIENCIAS (Colombia); MSES (Croatia); RPF (Cyprus); MoER, SF0690030s09 and ERDF (Estonia); Academy of Finland, MEC, and HIP (Finland); CEA and CNRS/IN2P3 (France); BMBF, DFG, and HGF (Germany);
GSRT (Greece); OTKA and NKTH (Hungary); DAE and DST (India); IPM (Iran); SFI (Ireland); INFN (Italy); NRF and WCU (Korea); LAS (Lithuania); CINVESTAV, CONACYT, SEP, and UASLP-FAI (Mexico); MSI (New Zealand); PAEC (Pakistan); MSHE and NSC (Poland); FCT (Portugal); JINR (Armenia, Belarus, Georgia, Ukraine, Uzbekistan); MON, RosAtom, RAS and RFBR (Russia);
MSTD (Serbia); MICINN and CPAN (Spain); Swiss Funding Agencies (Switzerland); NSC (Taipei); TUBITAK and TAEK (Turkey); STFC (United Kingdom); DOE and NSF (USA).

\vspace*{3ex} 
\bibliography{auto_generated}

\cleardoublepage \appendix\section{The CMS Collaboration \label{app:collab}}\begin{sloppypar}\hyphenpenalty=5000\widowpenalty=500\clubpenalty=5000\textbf{Yerevan Physics Institute,  Yerevan,  Armenia}\\*[0pt]
S.~Chatrchyan, V.~Khachatryan, A.M.~Sirunyan, A.~Tumasyan
\vskip\cmsinstskip
\textbf{Institut f\"{u}r Hochenergiephysik der OeAW,  Wien,  Austria}\\*[0pt]
W.~Adam, T.~Bergauer, M.~Dragicevic, J.~Er\"{o}, C.~Fabjan, M.~Friedl, R.~Fr\"{u}hwirth, V.M.~Ghete, J.~Hammer\cmsAuthorMark{1}, N.~H\"{o}rmann, J.~Hrubec, M.~Jeitler, W.~Kiesenhofer, M.~Krammer, D.~Liko, I.~Mikulec, M.~Pernicka$^{\textrm{\dag}}$, B.~Rahbaran, C.~Rohringer, H.~Rohringer, R.~Sch\"{o}fbeck, J.~Strauss, A.~Taurok, F.~Teischinger, P.~Wagner, W.~Waltenberger, G.~Walzel, E.~Widl, C.-E.~Wulz
% ...
\vskip\cmsinstskip
\dag:~Deceased\\
1:~~Also at CERN, European Organization for Nuclear Research, Geneva, Switzerland\\
% ...

\end{sloppypar}
\end{document}